\documentclass{article}

\usepackage{arxiv}

\usepackage[utf8]{inputenc} 
\usepackage[T1]{fontenc}    
\usepackage{hyperref}       
\usepackage{url}            
\usepackage{booktabs}       
\usepackage{amsfonts}       
\usepackage{nicefrac}       
\usepackage{microtype}      
\usepackage[subrefformat=parens]{subcaption}
\usepackage{lipsum}
 \usepackage{amsmath} 
\usepackage{graphicx}
\graphicspath{ {./images/} }

\title{Fast convolution kernels on pascal GPU with high memory efficiency}

\author{
 Qiong Chang \\
  Grad. of Systems and Information Engineering\\
  University of Tsukuba\\
  Tsukuba, Japan, 305-8577 \\
  \texttt{q.chang@c.titech.ac.jp} \\
   \And
  Masaki Onishi \\
  National Institute of Advanced Industrial Science and Technology (AIST)\\
  Tsukuba, Japan, 305-8560 \\
  \texttt{onishi@ni.aist.go.jp} \\
  \And
  Tsutomu Maruyama \\
  Grad. of Systems and Information Engineering\\
  University of Tsukuba\\
  Tsukuba, Japan, 305-8577 \\
  \texttt{maruyama@darwin.esys.tsukuba.ac.jp} \\
}

\begin{document}
\maketitle
\begin{abstract}
The convolution computation is widely used in many fields, especially in
CNNs.
Because of the rapid growth of the training data in CNNs, GPUs have been
used for the acceleration, and memory-efficient algorithms are focused
because of thier high performance.
In this paper, we propose two convolution kernels for single-channel
convolution and multi-channel convolution respectively. Our two methods
achieve high performance by hiding the access delay of the global memory
efficiently, and achieving high ratio of floating point Fused
Multiply-Add operations per fetched data from the global memory.
In comparison to the latest Cudnn library developed by Nvidia aimed to
accelerate the deep-learning computation, the average performance
improvement by our research is 2.6X for the single-channel, and 1.4X for
the multi-channel.

\end{abstract}


\section{Introduction}
Convolution is widely used as a fundamental operation in many
applications such as computer vision, natural language processing,
signal processing. Especially, the Convolution Neural Network (CNN), a
popular model used for deep-learning, is widely used in many
applications such as image recognition \cite{hand}\cite{VGG}, video
analysis\cite{Deep-CNN}, natural language processing\cite{NLP}, and has
yielded remarkable results.

Recently, many CNN models have been proposed, such as AlexNet
\cite{Alex}, GoogleNet \cite{GoogleNet}, VGGNet \cite{VGG}, ResNet
\cite{ResNet}, etc. They are used in many areas, and are improved
steadily.
The sizes of their networks have grown larger, and this
leads to the increase of their processing time.
The CNNs have several layers, but a large portion of the total
processing time is used for the convolution layers.
Because of the high inherent parallelism of the convolution algorithms,
many researchers are exploring to use GPUs to accelerate them.
They can be divided into four categories: 1) \textit{Direct-based}
method, 2) \textit{FFT-based} method, 3) \textit{Winograd-based} method
and 4) \textit{General matrix multiplication (GEMM) based} method.
%
%
%
%
\newpage

Recently, many of the algorithms and their modified versions have been
aggregated into a public library called Cudnn by Nvidia, which aims to
accelerate the deep-learning platforms like Chainer\cite{Chainer}, Caffe\cite{Caffe}, etc.
As the volume of the processing data used in deep-learning increases,
the memory-efficient algorithms play an increasingly more important
role, resulting in the constant proposal of many improved versions.
Among them, \textit{Implicit-GEMM}\cite{Cudnn} has been included in the
Cudnn because of its high efficiency.
It divides the feature map and the filter data into many sub-blocks and
converts them to many sub-matrices by using only on-chip memory of GPU,
not using global memory. This method is very memory-efficient, and
achieved a high ratio of floating point Fused Multiply-Add (FMA)
operations per data transferred from the global memory.
In \cite{Op-direct}, two memory-efficient methods were proposed.  Both
of them are faster than the Implicit-GEMM, however their performances
are negatively affected when the feature map size is smaller than 32,
because it fixes the amount of the data assigned to each $SM$, which
sometimes is not suitable to the small feature map. In CNN
models such as \cite{Alex}\cite{GoogleNet}\cite{VGG}\cite{ResNet}, more
than half of the convolution layers are used for the calculation of the
images smaller than 32 (such as 28, 14, 7). This means that
\cite{Op-direct} cannot handle the modern CNN models efficiently.

In this paper, we propose two methods to accelerate the convolution
kernels for single-channel and multi-channel respectively.  In these
methods, the memory access is optimized to achieve higher performance.
For the single-channel kernel, our approach follows the computation
method proposed in \cite{Op-direct}, however, the data are divided and
assigned to each $SM$ carefully to hide the access delay of the global
memory considering the input data size and the hardware features of our
target GPUs (Pascal GPUs).
%
%
For the multi-channel kernel, we propose a \textit{stride-fixed block}
method.
%
This method aims to maximize the number of FMA operations per loaded
data because the total amount of data that have to be loaded to each
$SM$ is much larger than the single-channel convolution, and the access
delay can be hidden by \textit{data prefetching}.
%

\section{Optimization Methods}
%
%
In this section, we first introduce the CNN models, and then discuss
what kind of optimization method is applicable on Pascal GPUs.

\subsection{The Convolution Models}
The multi-channel convolution can be defined as follows:
\begin{equation} 
  \begin{split}
    O^m(x,y) = \sum_{ch=1}^C\sum_{i=0}^{K-1}\sum_{j=0}^{K-1}I^{ch}(x+i,y+j)\cdot F^{ch,m}(i,j), \\
    where \hspace*{0.5cm} x\in[0,W_{x}-K+1), y\in[0,W_{y}-K+1), m\in[1,M].
  \end{split}
  \label{eq:abc}
\end{equation}
Here, $I$ is the input feature map set and $F$ is the input filter set. $O$ is
the output feature map set which is generated from $I$ and $F$. $x$ and $y$ are
the coordinates of the pixels of the feature maps. $W_{x}$ is the width and
$W_{y}$ is the height of the input feature map. $i$ and $j$ are the offsets, and
are added to the coordinates. Their upper bound $K$ decides the size of filter.
%
$ch$ represents the channel of the input in the range of $[1,C]$ ($C$ is
the number of channels), and all of the convolution results are
added along the dimension $ch$.
$m$ represents the filter number, and each filter has $C$ channels.
%
%
$M$ is the number of filters, and it is defined in each convolution
layer. 
%
%
When $C=1$, it is called single-channel convolution, and its
definition is given the following equation.
\begin{equation}
  \begin{split}
    O^m(x,y) = \sum_{i=0}^{K-1} \sum_{j=0}^{K-1}I(x+i,y+j) \cdot F^m(i,j)\\
  \end{split}
\end{equation}

\subsection{Acceleration models on GPU}
Here, we discuss the acceleration methods of the convolution
calculation.
However, this discussion is not restricted to the convolution, and can
be applied to other applications.

In GPUs, the on-chip memory including registers and shared memory, and
the off-chip memory, mainly the global memory are supported. The
register is the fastest, and the global memory is the slowest and
largest.
To fully utilize this hierarchy, many studies such as \cite{Op-direct}
have been proposed. However, throughout all computation, data loading
time from the global memory to the on-chip memory is most critical, and
hiding the latency of the global memory is the most important point for
the acceleration.

To hide the latency of the global memory, two methods can be considered:
\begin{enumerate}
\item
keep the operation units busy (mainly Fused Multiply-Add (FMA) operation units
in convolution) by executing more than $N_{FMA}$ operations (the lowest value to
make the units busy) in each $SM$ for the current data set until the next data
set arrive from the global memory by data prefetching, and
\item
transfer a large volume of data ($V_s$) from the global memory
continuously.
\end{enumerate}
In most cases, the first approach is preferable, because the data
loading overhead from the global memory can be relatively reduced more
by executing more number of operations per loaded data.
In the multi-channel convolution, the data size is large enough, and it
is possible to find the division of the feature maps and filters that
makes it possible to execute more than $N_{FMA}$ operations in each
$SM$.
However, in the single-channel convolution, when the size of feature
maps is small, the number of executable operations becomes less than
$N_{FMA}$ even with the data prefetching, and the second
approach is required.
Thus, it is necessary to make it clear under what conditions which
method shows better performance.

Table \ref{tb:freq} shows several parameters of GTX 1080Ti and its
performance for accessing single precision data.
As shown in Table \ref{tb:freq}, in GTX 1080Ti, 2
FMA operations can be executed in one clock
cycle in each core, namely 256 FMA operations in each $SM$ (each $SM$
has $N_{cores}$ = 128 cores).
According to the method proposed in \cite{Mem-hierarchy}, the global
memory latency of the GTX 1080Ti is 258 clock cycles.
In order to hide this 258 clock cycles, $N_{FMA} = 66,048$
FMA operations ($66,048 = 258 \times N_{cores} \times 2$) are
required in each $SM$ for the current data set (the set of divided
feature maps and filters).
%

The volume size $V_s$ can be calculated as follows.
The Geforce GTX 1080Ti has a base clock of 1480 MHz and the bandwidth
of 484 GB/s, which means the transfer rate is roughly 327 bytes per
clock cycle.
Therefore, the volume size which is needed to hide the latency (258
clock cycles) becomes 84,366 = 327 $\times$ 258 bytes.
To realize the data transfer of this size, 21,120 ($=$ 84,366 / 4)
threads are required because each thread fetches a 4 byte data in single
precision. Thus, in each of 28 $SMs$ ($N_{sm} = 28$ is the total number
of $SMs$ in the GTX 1080Ti), 768 threads are required to fetch one 4-byte
word respectively (in total, it becomes 768 $\times 4 \times 28 = $
86,016 > 84,366).
This means that the minimum volume size to make the global memory busy
is $V_s = $ 86,016 bytes.
For dividing the feature maps and filters, and assigning them to each
$SM$, the following procedure should be taken:
\begin{enumerate}
\item Divide the feature maps and filters so that the total size of data that
  are assigned to each $SM$ is smaller than the size of the
 shared memory $S_{shared}$ (96KB in GTX 1080Ti).
\item \label{lb:2nd} Evaluate the number of FMA operations that can be
  executed for the data in each $SM$.
\item If it is larger than $N_{FMA}$, use the the first method which is
  based on the data prefetching.
\item If not, redivide the feature maps and filters so that the total
  size of data that are transferred to all $SMs$ becomes larger than
  $V_s$, and use the second approach.
\end{enumerate}
Additionally, for accessing global memory, it is necessary to confirm
that the starting address and the size of the sequential accessing
segment is a multiple of \textit{32-byte}.
In Pascal GPU, a multiple of \textit{128-byte} shows better performance
than that of \textit{32-byte} and \textit{64-byte}, but the performance
for \textit{32-byte} and \textit{64-byte} is acceptable.
%

\begin{table}[t]
  \centering
  \caption{Parameters to access single precision data}
  \label{tb:freq}
   \begin{tabular}{r|c} 
    &GTX 1080Ti\\\hline
    Architecture &Pascal\\
    Global Memory Latency (clock cycles)&258\\
    Bandwidth (Gb/s)&484 \\
    Base clock cycle (MHz)&1480 \\
    SM&28 \\
    Transmission Rate (Byte/clock cycle) &327 \\
    Data Requirement (bytes) &84,366 \\
    Thread Requirement/SM &768 \\
    Warp Requirement/SM  &24 \\
    Data Requirement/SM (bytes)&3072\\
    Flops/clock cycle/core &2 \\
    \hline
\end{tabular}
  \vspace*{0.8cm}
\end{table}

\subsection{Data Mapping}

\begin{figure}[t]
  \begin{center}
    \begin{tabular}{c}
      \begin{minipage}{0.5\hsize}
        \begin{center}
          \includegraphics[clip, width=2.7in]{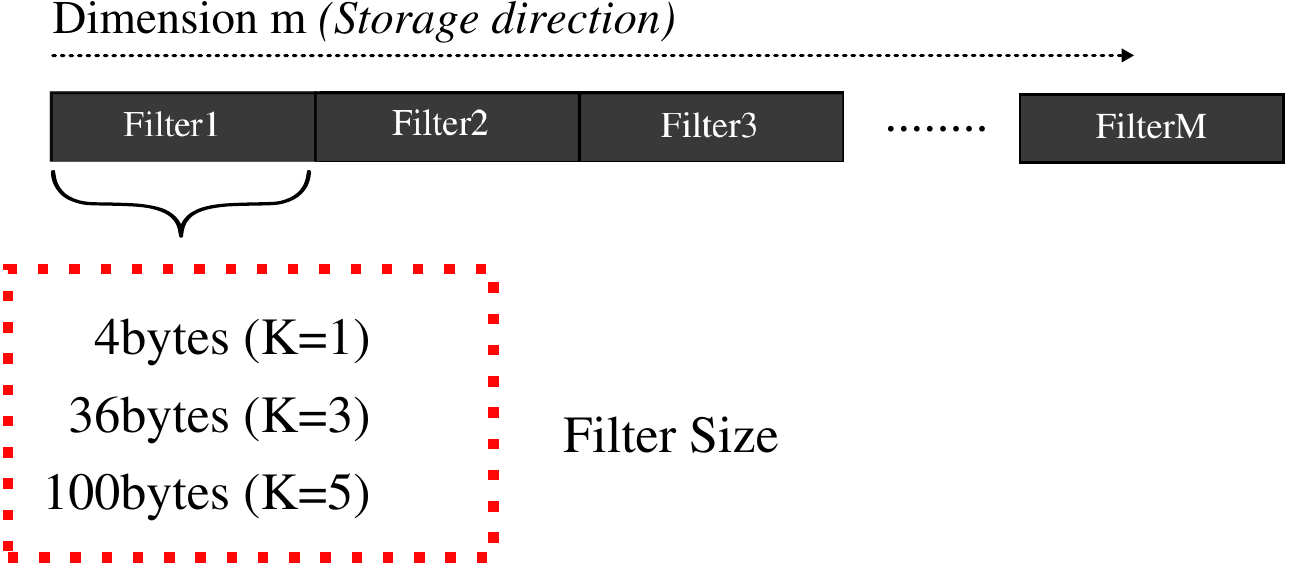}
          {\\(a) Single-Channel}
        \end{center}
      \end{minipage}
      \begin{minipage}{0.5\hsize}
        \begin{center}
          \includegraphics[clip, width=2.7in]{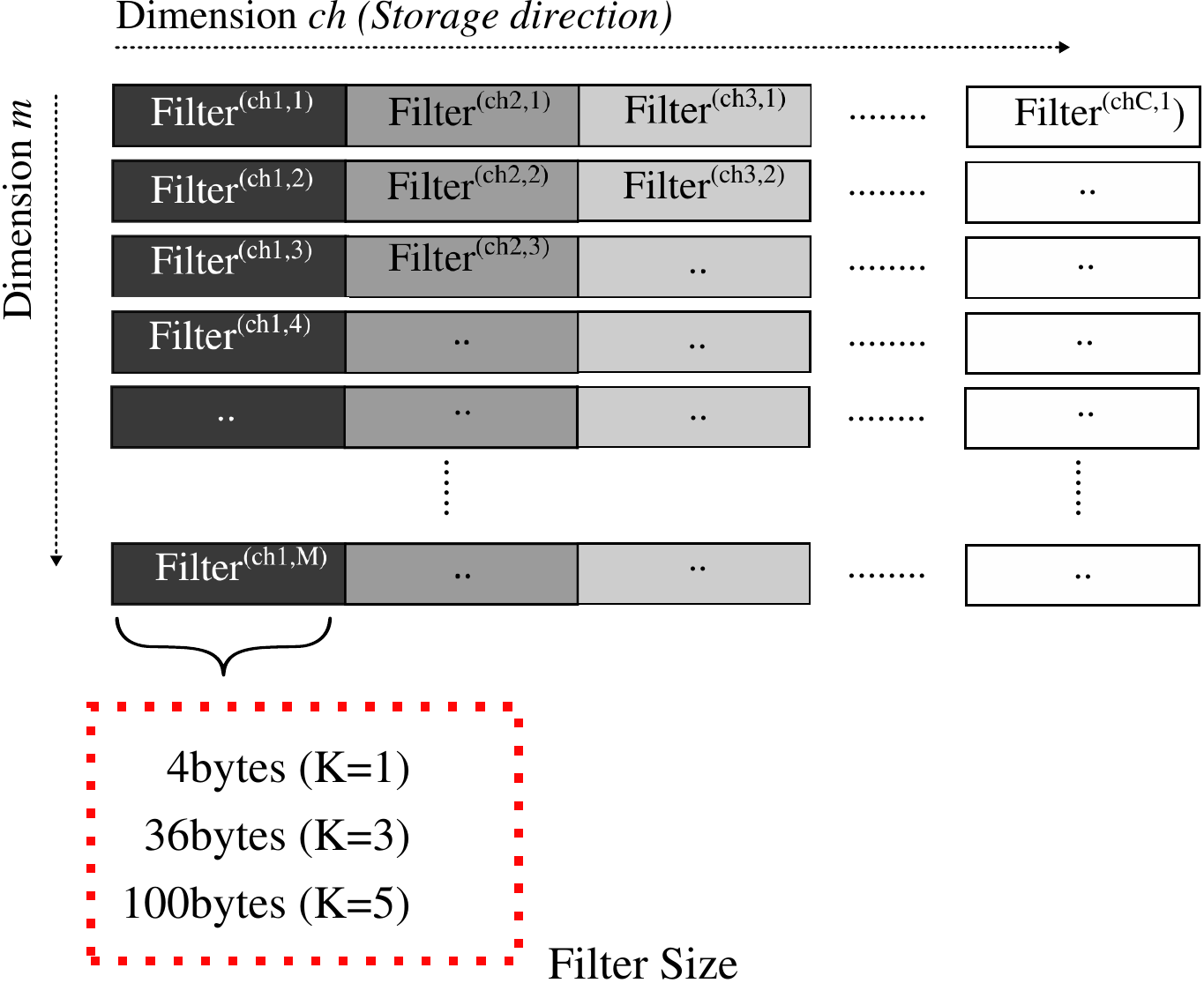}
          {\\(b) Multi-Channel}
        \end{center}
      \end{minipage}
    \end{tabular}
    \caption{Memory storage form of the Filter}
    \label{fig:mem_store}
  \end{center}
\end{figure}

As shown in Fig.\ref{fig:mem_store}(a), in the single-channel
convolution ($C=1$), the size of each filter is $K\times K
\times{4\mbox{-}byte}$, and they are stored in the global memory continuously.
With this data mapping, the filters can be divided only along the
dimension $m$, and the filters can be loaded from the global memory
efficiently because they are stored continuously.
Three approaches for dividing the feature maps and filters can be
considered.
\begin{enumerate}
\item 
Only the filters are divided. They are assigned to each $SM$, and in
each $SM$, the assigned filters are applied to the whole feature maps
(the feature map is processed sequentially against each filter).
\item 
Only the feature maps are divided. They are assigned to each $SM$, and in
each $SM$, the assigned feature maps are processed by all filters (the
filters are applied sequentially).
\item
Both feature maps and filters are divided, and they are assigned to each
$SM$ (the combination of the first and the second approach).
\end{enumerate}
By using different approach, the amount of data that has to be loaded to the
shared memory memory from the global memory, and the number of
FMA operations that can be executed in parallel become
different.
%
Therefore, finding a good balance between the size of divided feature
maps and filters becomes a key point.
%
%

\begin{figure}[t]
  \begin{center}
    \begin{tabular}{c}
      \begin{minipage}{0.45\hsize}
        \begin{center}
          \includegraphics[clip, width=2.7in]{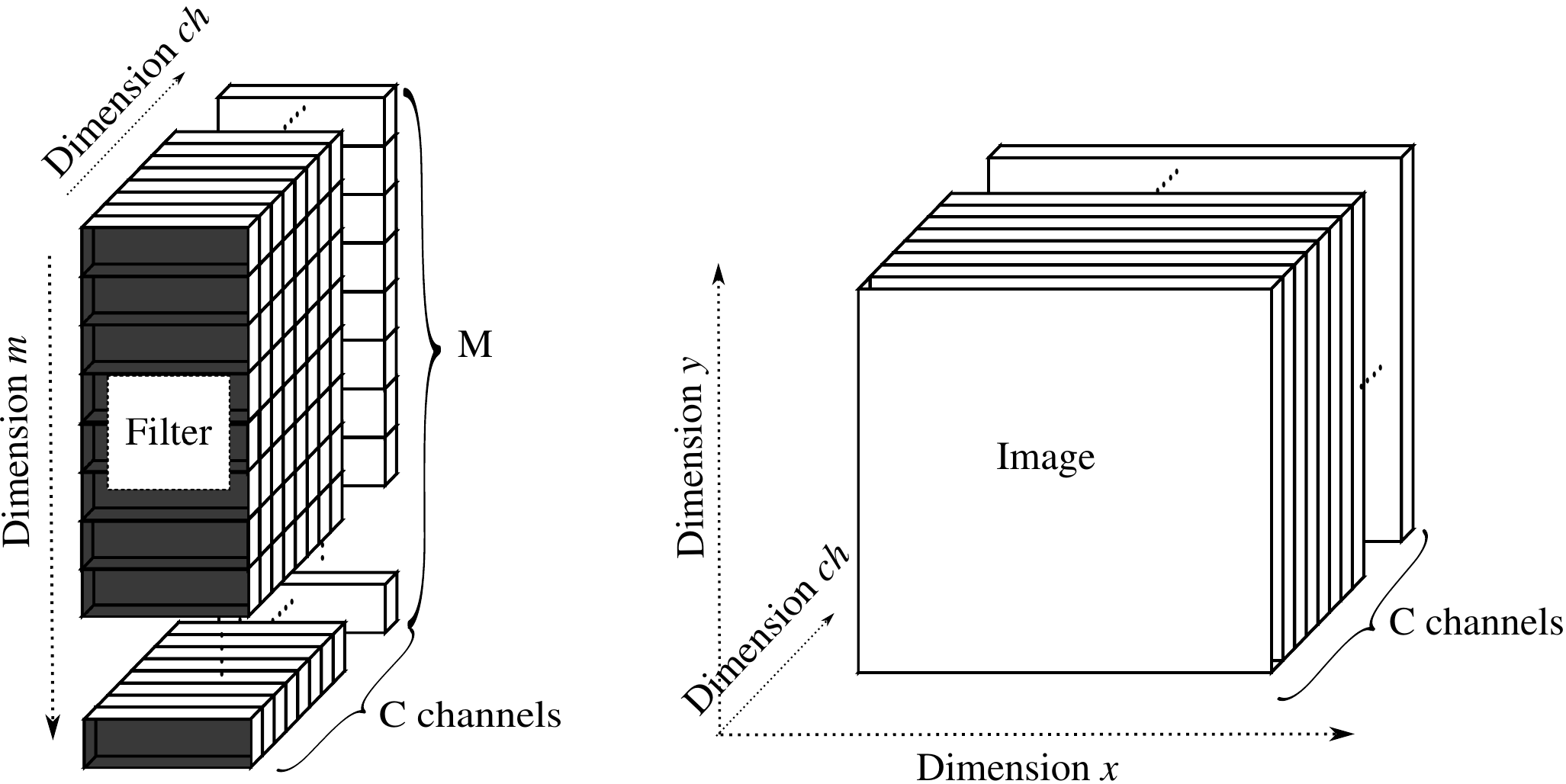}
          {\\(a)}
        \end{center}
      \end{minipage}
      \hspace*{0.5cm}
      \begin{minipage}{0.45\hsize}
        \begin{center}
          \includegraphics[clip, width=2.7in]{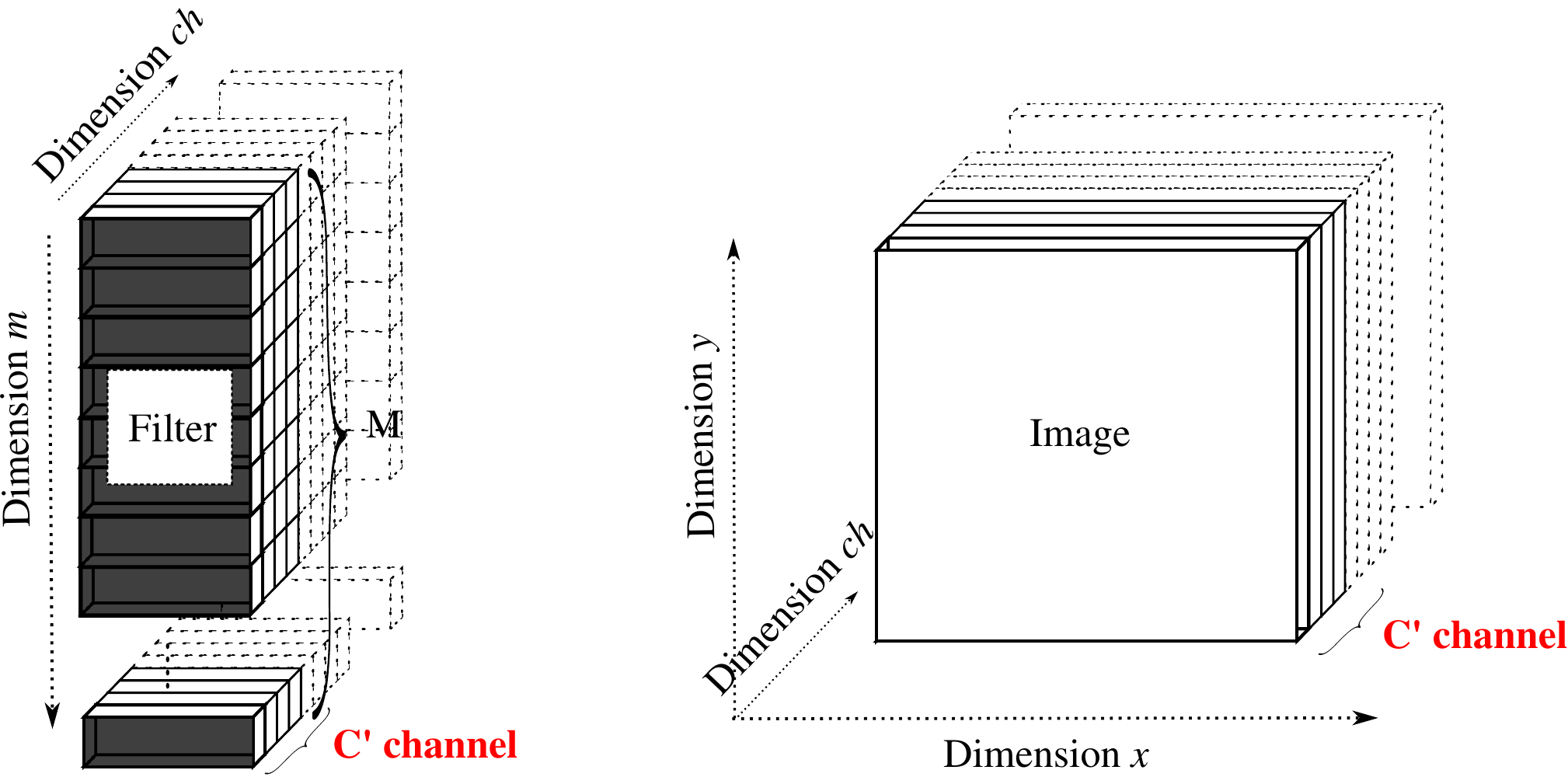}
          {\\(b)}
          \end{center}
      \end{minipage}
      \vspace*{0.1cm}
      \\
      \begin{minipage}{0.45\hsize}
        \begin{center}
          \includegraphics[clip, width=2.7in]{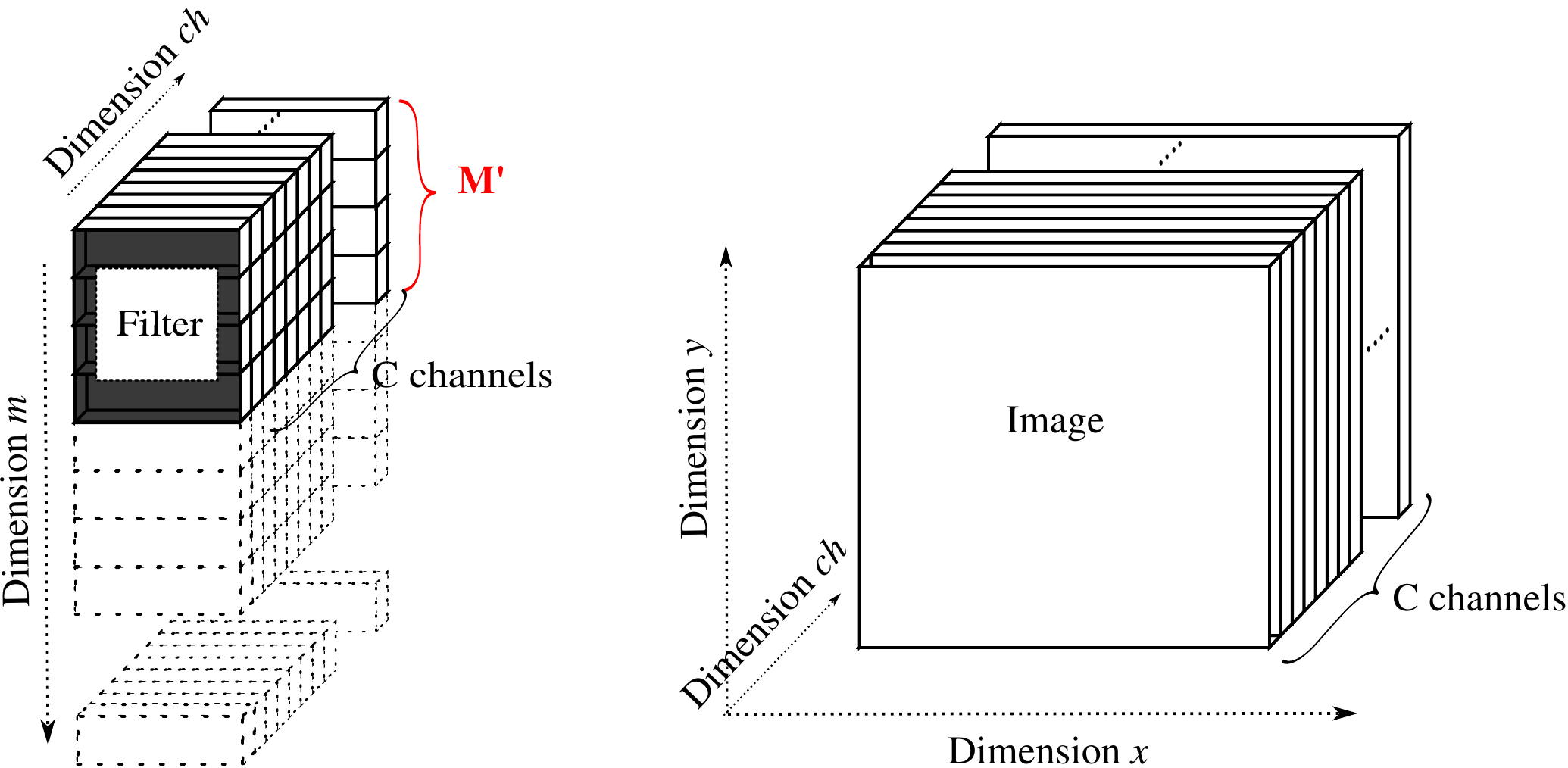}
         {\\(c)}
        \end{center}
      \end{minipage}
        \hspace*{0.5cm}
       \begin{minipage}{0.45\hsize}
        \begin{center}
          \includegraphics[clip, width=2.7in]{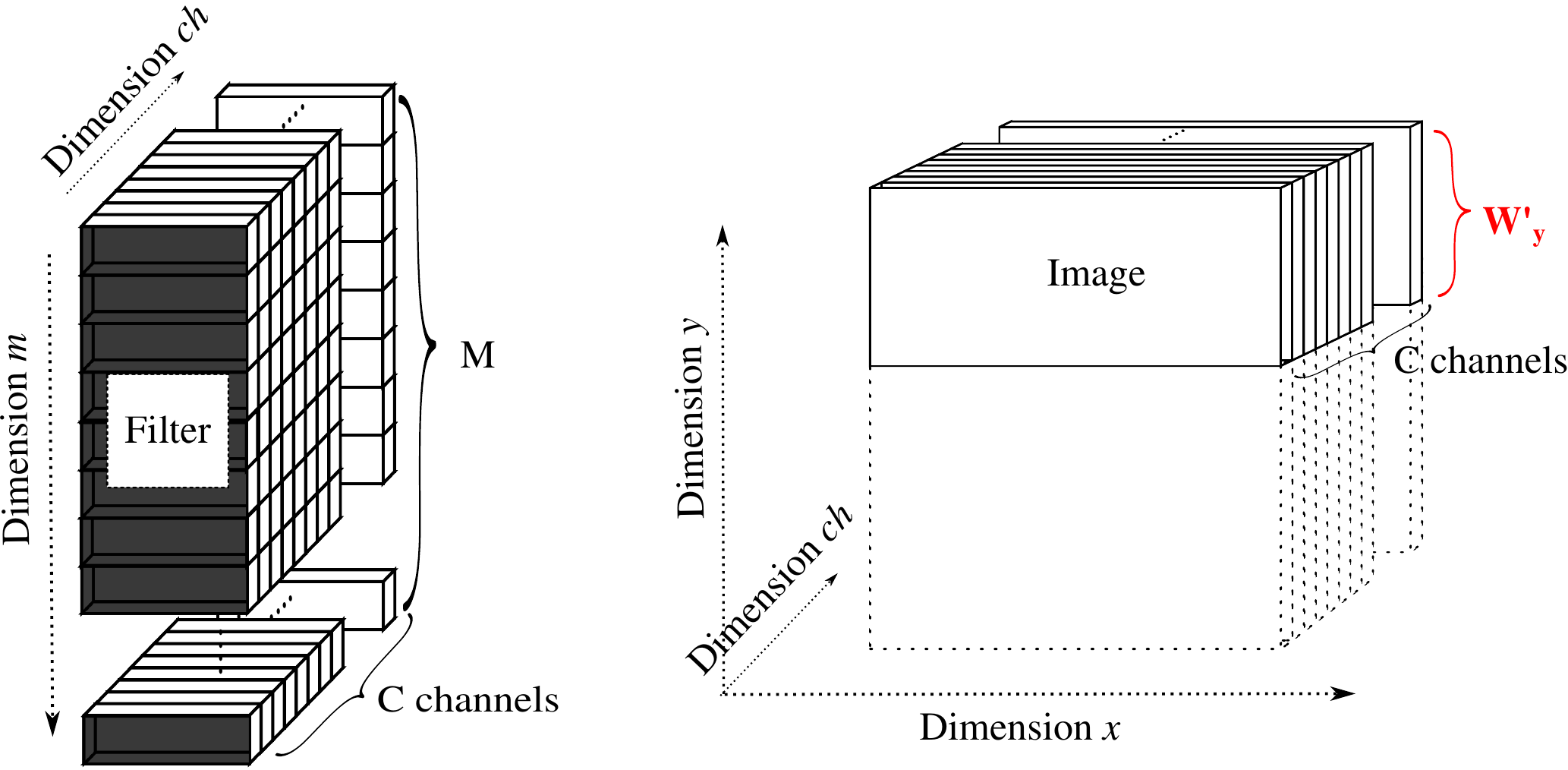}
         {\\(d)}
        \end{center}
       \end{minipage}
       \vspace*{0.1cm}
       \\
       \hspace*{-0.4cm}
        \begin{minipage}{0.45\hsize}
        \begin{center}
          \includegraphics[clip, width=2.7in]{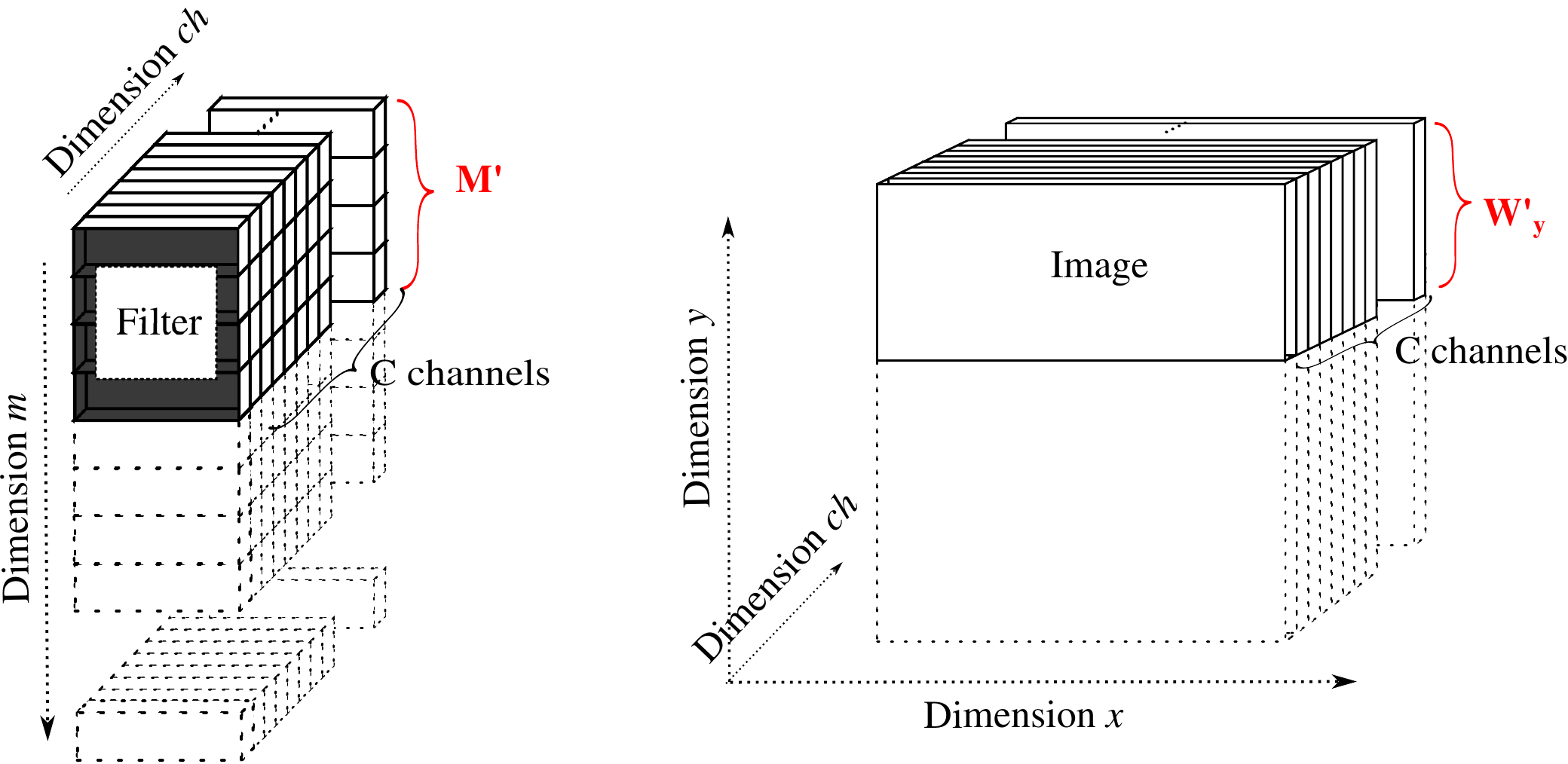}
         {\\(e)}
        \end{center}
        \end{minipage}
        \hspace*{0.5cm}
        \begin{minipage}{0.45\hsize}
        \hspace*{5.5cm}
        \end{minipage}
    \end{tabular}
    \caption{Assignment of the input data}
    \label{fig:assignment}
  \end{center}
\end{figure}

In the multi-channel convolution ($C>1$), which is the typical case in the
convolution layers of the CNN except for the first one, the data size becomes
much larger than the single-channel convolution.
%
%
%
Fig.\ref{fig:mem_store}(b) shows how the filters are stored in the
global memory.
They are stored along the dimension $ch$ first, and then along
the dimension $m$.
%
%
In this case, the dividing method of the filters along the dimension $m$
as used for the single-channel convolution can not be applied as it is,
because the data size of each filter is normally not a multiple of
\textit{32-byte}.
Especially when $K=1$ (the filter size is only 4 bytes), the filters are
accessed as \textit{4-byte} segments, and it causes serious performance
reduction because of \textit{non-coalescing memory access}.To solve this
problem, several approaches can be considered.
Fig.\ref{fig:assignment}(a) shows the whole data structure before the
data division.
\begin{enumerate}
\item   
In Fig.\ref{fig:assignment}(b), both the filters and the feature maps
are divided along the dimension $ch$, and the data for $C' = C/N_{sm}$
channels are assigned to each $SM$ ($N_{sm}$ is the total number of
$SMs$).
%
%
With this division, the data calculated in each thread have to be summed
along dimension $ch$.
This means that add operations among $SMs$ are required, and
$W_{x}\times{W_{y}\times{C'}\times{4\mbox{-}byte}}$ byte in the global
memory are used for this summation.
The global memory accesses to this area and the synchronous operations
required for this summation considerably reduce the overall performance.
%
\item 
In Fig.\ref{fig:assignment}(c), only the filters are divided along the
dimension $m$. $M' \times C = M/N_{sm}\times C$ filters are assigned to
each $SM$, and the whole feature map are loaded to $SMs$ from the
global memory.
In this approach,
if the total size of the filters is less than the total size of all
shared memory ($ K \times K \times C \times M \times 4\mbox{-}byte <
N_{sm} \times S_{shared}$), the divided filters can be cached in each
$SM$, and no additional access to the global memory is required.
\item 
On the other hand, in Fig.\ref{fig:assignment}(d), only the feature maps
are divided along dimension $y$. The divided feature maps are assigned
to each $SM$, and the whole filters are loaded to $SMs$ from the global
memory.
In this case, if the total size of the feature maps is smaller than the
total size of the shared memory, the divided feature maps can be cached
in each $SM$, and no additional access to the global memory is required.
\item
However, the total size of the filters and feature maps are larger than
the total size of the shared memory in general.
Thus, as shown in Fig.\ref{fig:assignment}(e), both the filters and
feature maps have to be divided respectively, and each divided segments
is cached in each $SM$ or loaded from the global memory.
In this case, there exist many alternatives for how to divide the
filters and features maps.
\end{enumerate}
According to our preliminary evaluation, the performance with the data
dividing method along the dimension $ch$ (Fig.\ref{fig:assignment}(b))
is obviously slower than other dividing methods because of the
additional access to the global memory for the addition.
For achieving higher performance, it is necessary to choose other dividing
methods considering the data size and the hardware features of the target GPUs
so that in each $SM$ the number of FMA
operations that can be executed per loaded data from the global memory is
maximized.


\section{GPU Implementation}


According to the discussion in Section 2, in both the single-channel and
multi-channel convolution, it is important to make the number of
FMA operations per pre-fetched data higher than $N_{FMA}$ in
order to achieve higher performance.
However, in some cases of the single-channel convolution, for example
when the size of feature maps is small, the number of FMA
operations cannot be kept high enough by data prefetching.
This means that, in case of single channel convolution, according to the
size of input data, we need to choose one of the two methods described
in Section 2.2: data prefetching or data transfer larger than
$V_s$.

In the multi-channel convolution, the size of input data is large
enough, and the number of FMA operations can be kept high
enough by data prefetching.
However, the performance can be improved more by achieving higher
FMA operation ratio for the fetched data, because to fetch
the data from the global memory, each thread has to issue the
instruction to read data, and the clock cycles are spent for issuing
these read instructions.
Therefore, in the multi-channel convolution, to find the data dividing
method that maximizes the number of FMA operations for each
divided data is the key to achieve higher performance.

%
%

\subsection{Single-Channel Convolution}
%
%
Here, we describe how to divide the input data to achieve higher
performance in the single channel convolution.
As for the convolution calculation in each $SM$, we follow the method proposed in \cite{Op-direct}.

As shown in equation (4), the total amount of the input data is given
by:
\begin{equation} 
\begin{split}
 D_{input}=D_{filter}+D_{map}=(K\times{K}\times{M}+W_{x}\times{W_{y}})\times{4}\,Bytes.
\end{split}
\end{equation}
Let $N_{sm}$ be the number of $SMs$.
There are two ways to divide the input data and assign them to each
$SM$.
In the first method, the input data is divided along the dimension $m$
of the filter.
$D_{1}$, the size of input data assigned to each $SM$, becomes
\begin{equation}
\begin{split}
  D_{1}=\frac{D_{filter}}{N_{sm}}+D_{map}=(K\times{K}\times{\lceil\frac{M}{N_{sm}}\rceil}+W_{x}\times{W_{y}})\times{4}\,Bytes
\end{split}
\end{equation}
In general, $D_{map}$ is too large to be stored in the on-chip memory of
each $SM$. Thus, $D_{map}$ is divided into $P$ pieces along the
dimension $y$.
%
The size of each piece becomes $D_{Inc1} = D_{map} / {P}$.
Here, for each line of feature map, since the convolution requires
additional $K-1$ lines, the amount of data that have to be held in the
on-chip memory becomes
\begin{equation}
\begin{split}
  D_{1}=\frac{D_{filter}}{N_{sm}}+D_{Inc1}+(K-1)*W_{x}=(K\times{K}\times{\lceil\frac{M}{N_{sm}}\rceil}+(\lceil\frac{W_{y}}{P}\rceil+K-1)\times{W_{x}})\times{4}\,Bytes
\end{split}
\end{equation}
and the number of FMA operation that can be executed for
these data in each $SM$ is given by
\begin{equation}
\begin{split}
Th_{1}=\frac{D_{filter}}{N_{sm}}\times{D_{Inc1}}=K\times{K}\times{\lceil\frac{M}{N_{sm}}\rceil}\times{\lceil\frac{W_{y}}{P}\rceil\times{W_{x}}}.
\end{split}
\end{equation}

In the second method, the input data is divided along the dimension $y$
of the feature map.
In this case, $D_{2}$, the amount of the input data assigned to each
$SM$, becomes
\begin{equation}
\begin{split}
 D_{2}=D_{filter}+\frac{D_{map}}{N_{sm}}=(K\times{K}\times{M}+({\lceil\frac{W_{y}}{N_{sm}}\rceil+K-1})\times{W_{x}})\times{4}\,Bytes.
\end{split}
\end{equation}
$D_{filter}$ is too large to be stored in the on-chip memory in general,
and it is divided into $Q$ pieces.
The size of each piece becomes $D_{Inc2} = D_{filter} / {Q}$.
Then, $D_{2}$ becomes
\begin{equation}
\begin{split}
  D_{2}=D_{Inc2}+\frac{D_{map}}{N_{sm}}=\frac{D_{filter}}{Q}+\frac{D_{map}}{N_{sm}}=(K\times{K}\times{\lceil\frac{M}{Q}\rceil}+(\lceil\frac{W_{y}}{N_{sm}}\rceil+K-1)\times{W_{x}})\times{4}\,Bytes
\end{split}
\end{equation}
and the number of FMA operation that can be executed for
these data in each $SM$ is given by
\begin{equation}
\begin{split}
 Th_{2}=D_{Inc2}\times\frac{D_{map}}{N_{sm}}=K\times{K}\times{\lceil\frac{M}{Q}\rceil}\times({\lceil\frac{W_{y}}{N_{sm}}\rceil})\times{W_{x}}.
\end{split}
\end{equation}

The values of $P$ and $Q$ are decided considering if $D_{1}$ or $D_{2}$
is smaller than $S_{shared}$, and if $Th_{1}$ or $Th_{2}$ is larger than
$N_{FMA}$.
If $P=1$ or $Q=1$, the feature maps or the filters are not divided, and
they are transferred to the on-chip memory at a time.
If $P>1$ or $Q>1$, the feature maps or the filters are divided into
several pieces, and the pieces are transferred to each $SM$ by using the
data prefetching.
%
%
With smaller $P$ and $Q$, $D_{1}$, $D_{2}$ and $Th_{1}$, $Th_{2}$ become
larger.
The lower bound of $P$ and $Q$ is given by the requirement that $D_{1}$
and $D_{2}$ have to be smaller than $S_{shared}$, and the upper bound is
given by the requirement that $Th_{1}$ and $Th_{2}$ should be larger
than $N_{FMA}$.
$P$ and $Q$ should be chosen so that these requirements can be
satisfied.
In our implementation, $P$ and $Q$ are decided as follows.
\begin{enumerate}
\item $Th_{1}$ or $Th_{2}$ should be larger than the number of FMA operation $N_{FMA}$.\vspace*{0.1cm}\\
  \hspace*{0.5cm}$Th_{1} \ge  N_{FMA}$ and $Th_{2} \ge  N_{FMA}$\vspace*{0.1cm}\\
  Thus, the upper bound of $P$ and $Q$ (they must be smaller than $W_y$ and $M$ respectively) is given as follows \vspace*{0.15cm}\\
  \hspace*{0.2cm}$P\le\frac{K\times{K}\times{\lceil\frac{M}{N_{sm}}\rceil}\times{W_{y}}\times{W_{x}}}{N_{FMA}}
  \,and\,P\le{W_{y}}$\vspace*{0.15cm},\hspace*{1cm}$Q\le\frac{K\times{K}\times{M}\times{\lceil\frac{W_{y}}{N_{sm}}\rceil}\times{W_{x}}}{N_{FMA}}
  \,and\,Q\le{M}$\vspace*{0.15cm}
\item $D_{1}$ and $D_{2}$ must be smaller than the size of on-chip
  memory. The lower bound of $P$ and $Q$ is given as follows.\vspace*{0.15cm}\\
  \hspace*{0.3cm}$ P \ge\frac{4\times{W_{y}}\times{W_{x}}}{S_{shared}-4\times{K\times{K\times{\lceil\frac{M}{N_{SM}}}\rceil}}+(1-K)\times{4}\times{W_{x}}}$,
  \hspace*{0.5cm}$ Q \ge\frac{4\times{M}\times{K}\times{K}}{S_{shared}-4\times{W_{x}}\times{(\lceil\frac{W_{y}}{N_{SM}}+K-1)}}$\vspace*{0.15cm}\\
  Actually, there exist one more requirement to decide this lower bound.
  The number of required registers for the computation must be smaller
  than that supported in each $SM$. Its detail is not shown here, but
  considering this requirement, the lower bound is calculated.
\item If there exist $P$ and $Q$ ($P$ and $Q$ must be an integer) in the
  range specified by (2) and (3), Any of them can be used.  In our
  current implementation, the minimum ones are chosen as $P$ and $Q$,
  because the smaller values means less number of division, and make the
  processing sequence simpler. If no value exists, $P$ and $Q$ are set
  to 1.
\item Using the obtained $P$ and $Q$, $D_{1}$ and $D_{2}$ are calculated
  and compared. If $D_{1}$ is smaller than $D_{2}$, $Q$ is reset to 1
  to use the first dividing method described above, and otherwise, $P$
  is reset to 1 to use the second one. Both methods can be used because
  they both satisfy the requirements, but for the safety (for leaving
  more memory space on the on-chip memory), the smaller one is chosen.
\end{enumerate}
Following this procedure, the input data are divided and allocated to
each $SM$ in the best balance.
%
%
%

\subsection{Multi-Channel Convolution}
\begin{figure}[t]
\centering
\includegraphics[width=0.84\textwidth]{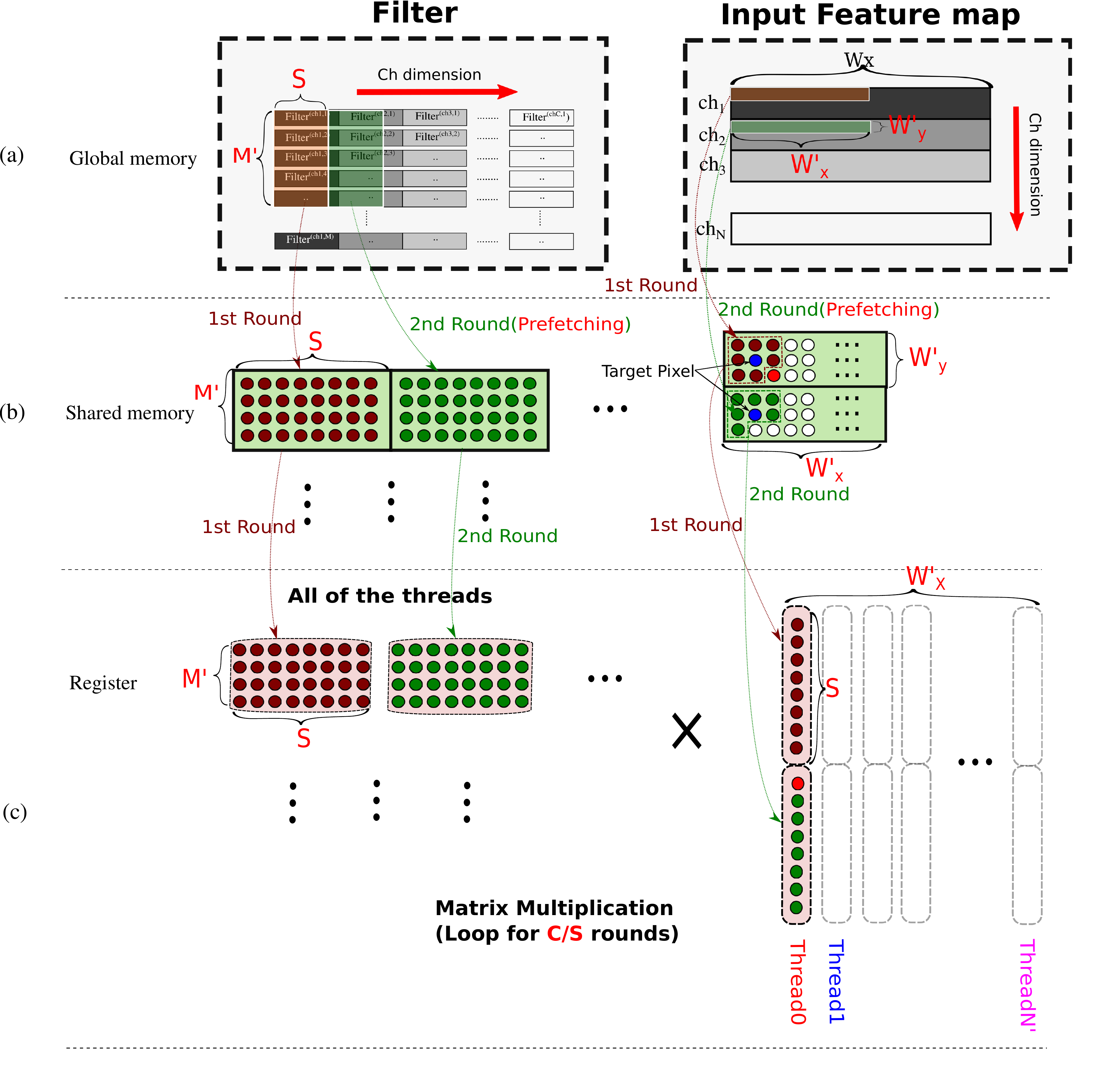}
\caption{Multi-Channel Convolution Kernel}
\label{fig:multi}
\end{figure}

As described above, in the multi-channel convolution, both feature maps
and filters are divided, and prefetching is used to transfer them to
each $SM$ from the global memory.
Recently, the block-based methods show high performance in convolution
due to their continuous and simple memory access sequence.
Fig.\ref{fig:multi} shows the data mapping of the filters and feature
map, and how they are divided and calculated in each $SM$.
In the block-based method, as shown in Fig.\ref{fig:multi}(a), the
following data are loaded to the on-chip memory in each $SM$.
\begin{enumerate}
\item
$S$ bytes of each filter along the dimension $ch$ (called
  \textit{segment} in the following discussion) of $M'$ filters
  ($S\times{M'}$ bytes in total), and
\item
a part of feature map, $W'_{x}\times{W'_{y}}\times {4\,bytes}$ in the
same channel ($W'_x$ is an arbitrary value that is decided by the size of
on-chip memory, but $W'_y$ is specified as
$\lceil\frac{S}{K\times{4bytes}}\rceil$, because when $S$ bytes are
fetched along the dimension $ch$,
$\lceil\frac{S}{K\times{4bytes}}\rceil$ lines in the feature map are
required to apply the filter).
\end{enumerate}

%
%
Then, the convolution is calculated for these data, and the next data
(next $S \times M'$ bytes of filters and $W'_{y}\times W'_{x}$ bytes of
feature maps are loaded by data prefetching.
%
%
%
In \cite{Op-direct}, the filter size is chosen as $S$ ($S =
K\times{K}\times{4\,bytes}$), and
only the filters of the target channel and a part of feature map of the
same $channel$ are loaded to the on-chip memory.
However, the filter size $K \times K $ is usually odd and often small, and
the performance is seriously degraded because of \textit{non-coalescing
  memory access}.
\cite{Dgemm} tried to solve this problem by extending $S$ to
\textit{128-bytes}.
By fetching continuous 128 bytes on the global memory, the highest
memory throughput can be achieved in GPUs.
In this method, the filters of several channels (and a part of the next
channel) are fetched at the same time, and they are kept in the on-chip
memory.
First, only the filters of the first channel are used for the
computation, and then, the filters of the next channel are used.
With this larger $S$, $M'$ has to be kept small because of the limited
size of on-chip memory, and smaller $M'$ means less parallelism ($M'$
filters are applied in parallel to the feature map of the same channel).
In \cite{Op-direct}, higher parallelism comes first, while in
\cite{Dgemm}, lower access delay has a higher priority.

Here, we propose a \textit{stride-fixed block} method not only to
maintain the efficient global memory access, but also to achieve high
parallelism in each $SM$.
\begin{enumerate}
\item
$S$ is set to a multiple of 32-bytes.
Actually, 32 or 64 is used.
Small $S$ allows larger $M'$, namely higher parallelism, under the
limited size of on-chip memory.
When $S$ is 32 or 64 bytes, the memory throughput from the global memory
becomes a bit worse than $S = 128$ bytes (the highest throughput), but
it is acceptable.
$S=32$ is the minimum value to maintain efficient global memory access.
\item
Next, we fix $W'_{x}$.
$W'_{x}$ pixels in the feature map are fetched along the dimension $x$
from the global memory.
Thus, $W'_{x}$ should be a multiple of 128-bytes to achieve the highest
memory throughput.
Larger $W'_{x}$ is preferable because it increases the Instruction Level
Parallelism (ILP), which can improve the performance of the convolution.
\item
After deciding the values of $S$ and $W'_{x}$, the most suitable $M'$
can be found by the requirements of the number of FMA
operations\vspace*{0.15cm}.\\
\hspace*{0.2cm}$M'\ge\frac{N_{FMA}\times{4\mbox{-}bytes}}{S\times{W'_{x}}}$\vspace*{0.15cm}.
\item
Because the data prefetching is used to fetch the next data set while
the current data set is being used for the current calculation, the size
of data set cannot exceeds the half of the size of shared memory.
Thus, \vspace*{0.15cm}\hspace*{0.2cm}$(S\times{M'}+\lceil\frac{S}{K\times{4bytes}}\rceil\times{W'_{x}})\le\frac{S_{shared}}{2}$\vspace*{0.15cm}\\
Here, $\lceil\frac{S}{K\times{4bytes}}\rceil = W'_y$ is the number of feature
maps required for the calculation.
\end{enumerate}
With this approach, for given $S$, $W'_x$ and $M'$ to achieve high
performance based on \textit{block method} can be obtained.

From here, we describe how the convolution calculation is executed in
each $SM$.
As shown in Fig.\ref{fig:multi}(a)(b), first, each $SM$ loads $S\,bytes$
of $M'$ filters to the shared memory.
At the same time, 
$W'_{x}$ pixels on $\lceil\frac{S}{K\times{4bytes}}\rceil = W'_y$ lines of the
feature maps are also loaded.
After the first round loading of these data, the same size of data for the
next round are pre-fetched: the next $S \times M'$ bytes along the
dimension $ch$, and the next $W'_x$ pixels of the $W'_y$ lines.
During the second round loading, the convolutions for the first round
data set are calculated on the chip as shown in
Fig.\ref{fig:multi}(b)(c).
On the chip, each thread corresponds to one target pixel of the feature
map as shown in Fig.\ref{fig:multi}(b).
Because the accessing speed of registers is faster than that of shared
memory, it is required to transfer each data in the shared memory to the
registers in order to achieve high performance.
In the convolution computation, the target pixel and its neighbors in
the feature map are sent to the corresponding registers by each thread.
Here, one important point is that only $S / {4bytes}$ pixels in the
feature map have to be loaded onto the registers.
The rest pixels, the red ones in Fig.\ref{fig:multi}(b), are just held
in the shared memory for the next round.
The filter data are also transferred to the registers by the
corresponding thread, but in this case, all data are transferred to the
registers because all of them are used.
After that, each pixel data is multiplied by the corresponding filter
data, and their products are added.
When all computations for the data stored in on-chip memory has been
finished, data prefetching for the third round is started.
During this loading, the convolution for the second round data is
calculated.

By using this method, the size of $S$ can be kept small, and the number
of filters $M'$ can be increased.
This ensures that more filters can be applied in parallel to the same
feature map.
This does not increase the number of data loading of the feature maps,
and hides the latency caused by global memory access.

\section{Experimental analysis}

We implemented our two convolution kernels on Pascal
series GPU Geforce GTX 1080Ti by using the CUDA 8.0.
Their performances were evaluated using many convolutions which
are commonly used in popular CNN
models\cite{Alex}\cite{ResNet}\cite{VGG}\cite{GoogleNet}, and compared with the
latest public library Cudnn v7.1\cite{Cudnn}.

\begin{figure}[t]
  \begin{center}
    \begin{tabular}{c}
      \begin{minipage}{0.33\hsize}
        \begin{center}
          \includegraphics[clip, width=2.3in]{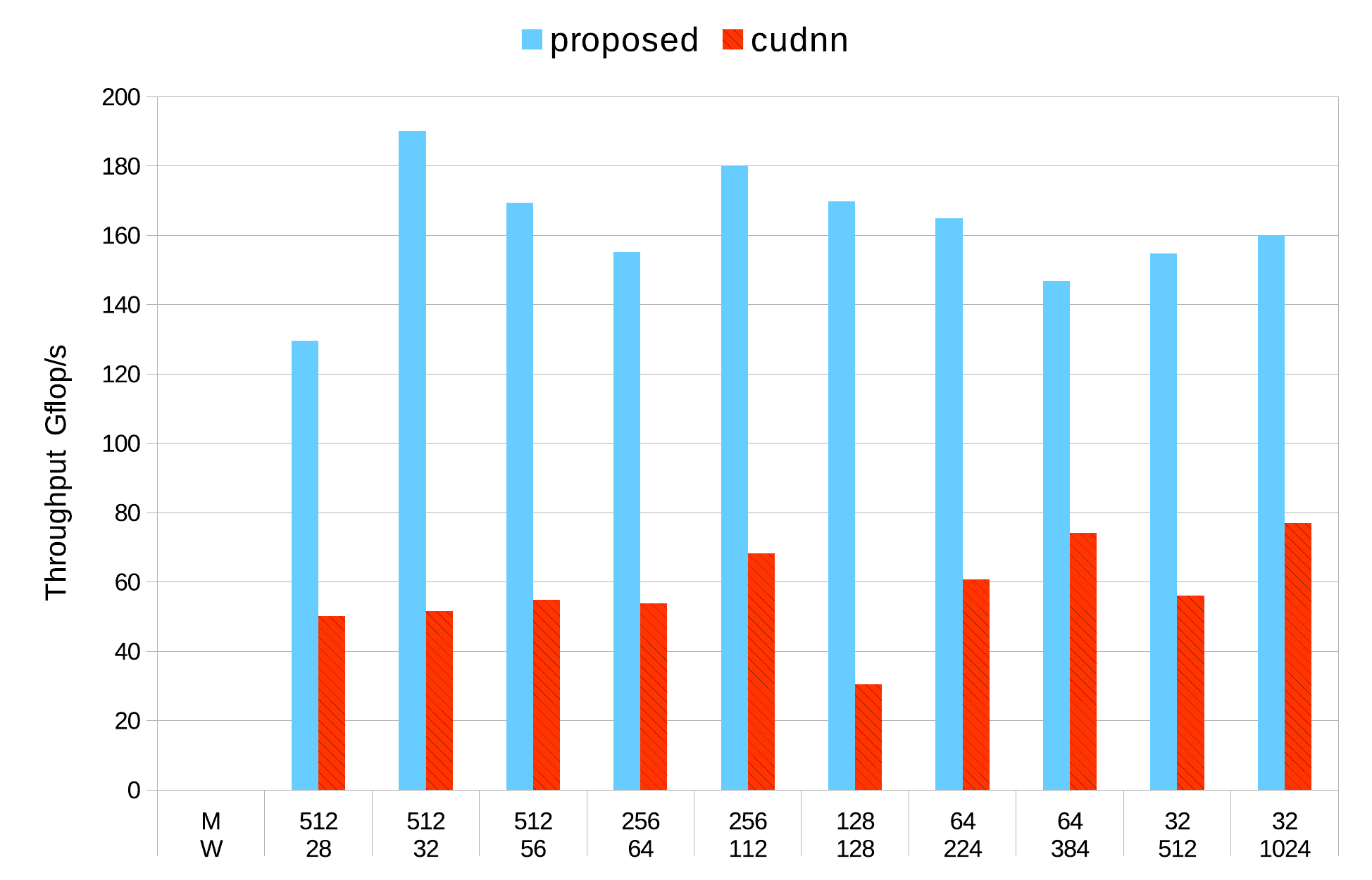}
          {Filter Size = 1}
        \end{center}
      \end{minipage}
      \begin{minipage}{0.33\hsize}
        \begin{center}
          \includegraphics[clip, width=2.3in]{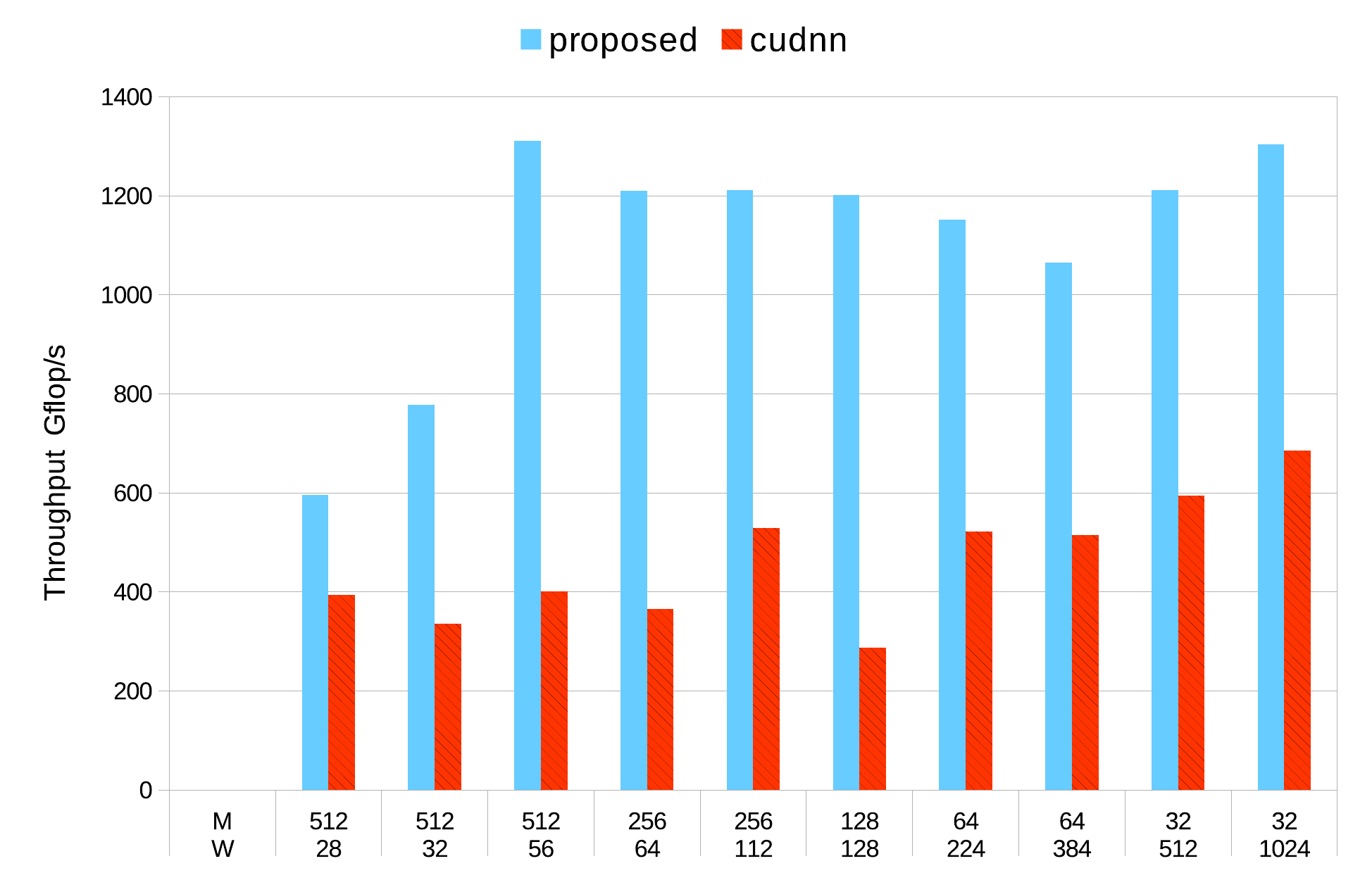}
          {Filter Size = 3}
        \end{center}
      \end{minipage}
      \begin{minipage}{0.33\hsize}
        \begin{center}
          \includegraphics[clip, width=2.3in]{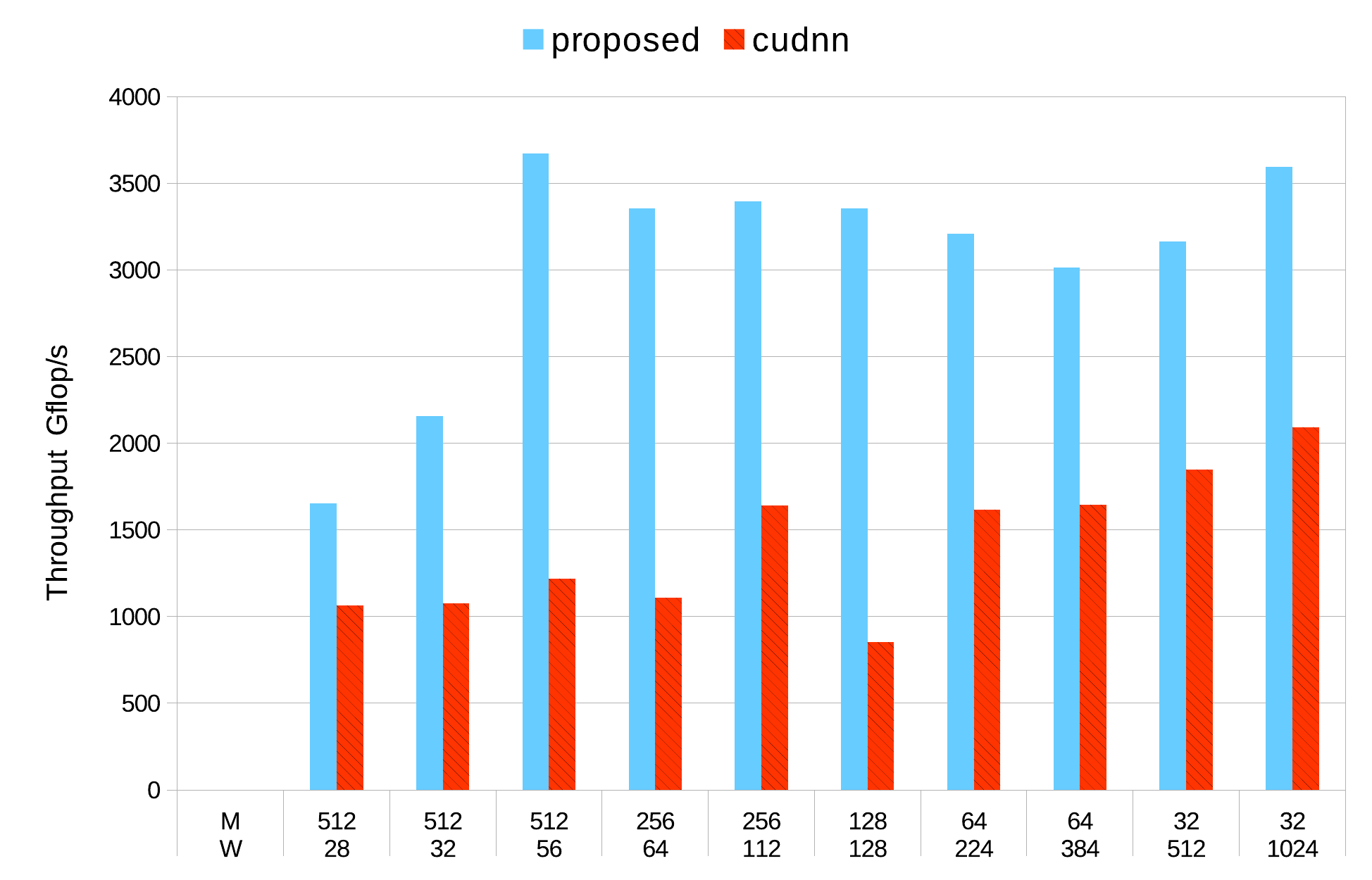}
         {Filter Size = 5}
        \end{center}
      \end{minipage}
    \end{tabular}
    \caption{Performance of the Single-Channel Convolution Kernel }
    \label{fig:single-result}
  \end{center}
  \begin{center}
    \begin{tabular}{c}
      \begin{minipage}{0.33\hsize}
        \begin{center}
          \includegraphics[clip, width=2.3in]{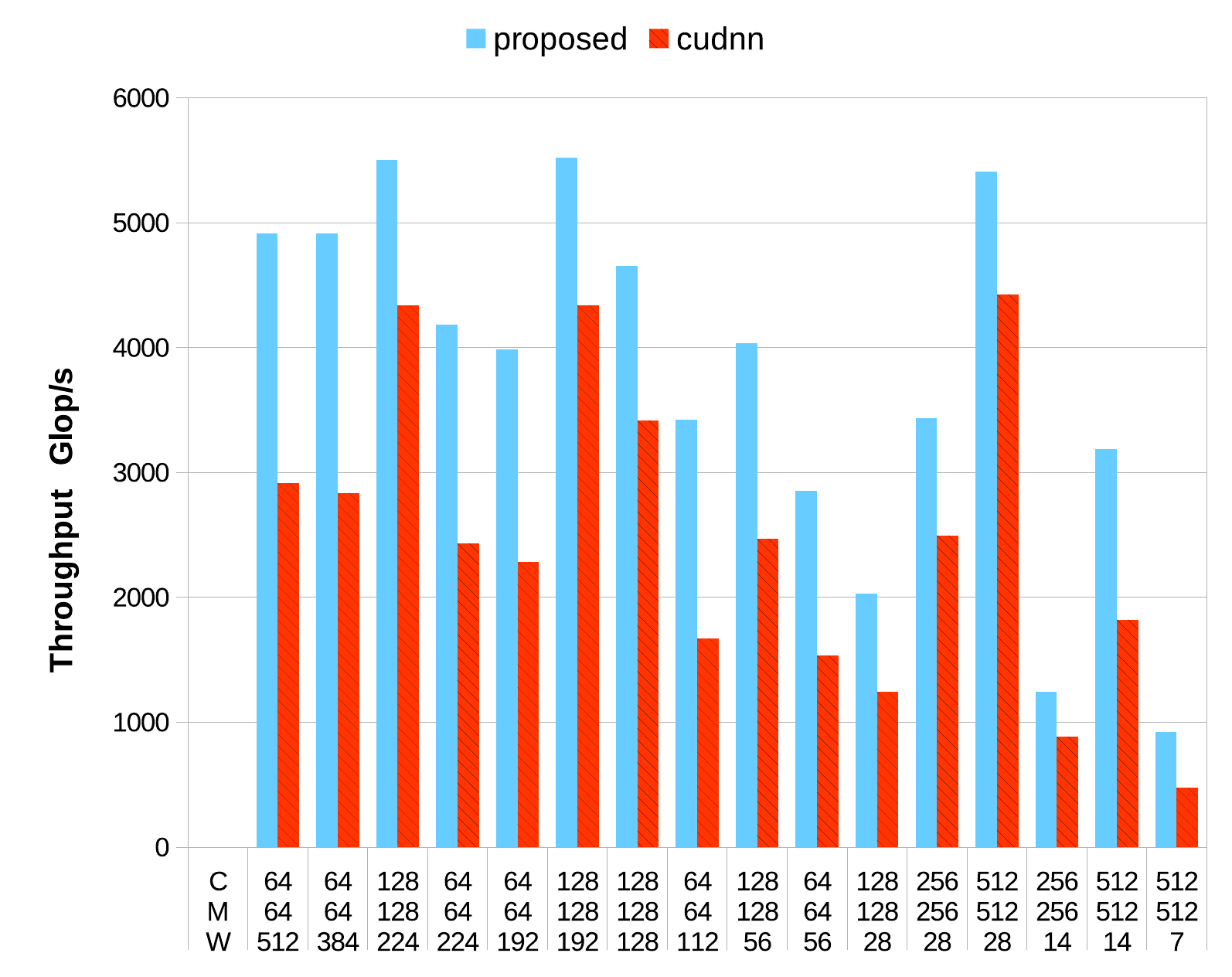}
          {Filter Size = 1}
        \end{center}
      \end{minipage}
      \begin{minipage}{0.33\hsize}
        \begin{center}
          \includegraphics[clip, width=2.3in]{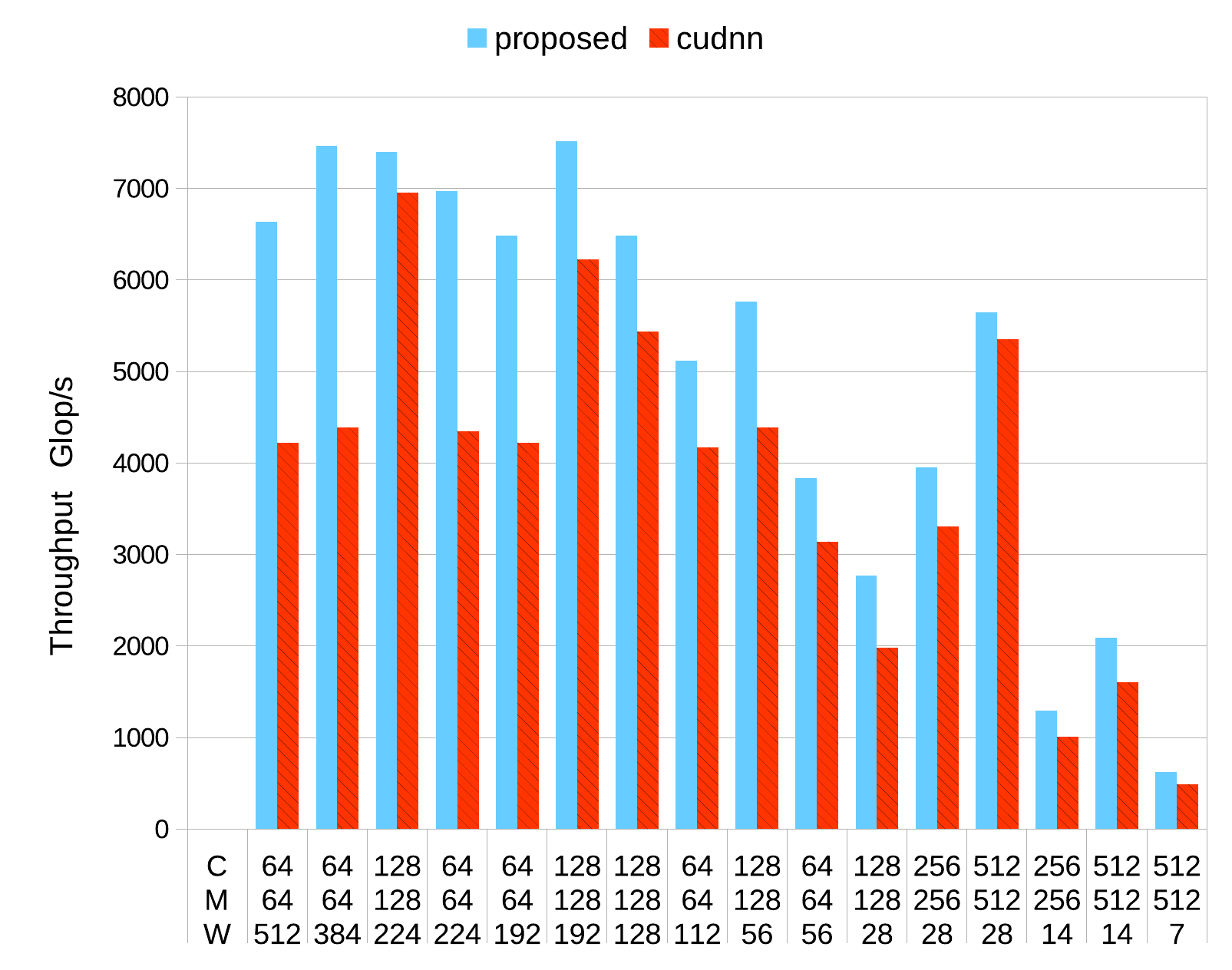}
          {Filter Size = 3}
        \end{center}
      \end{minipage}
      \begin{minipage}{0.33\hsize}
        \begin{center}
          \includegraphics[clip, width=2.3in]{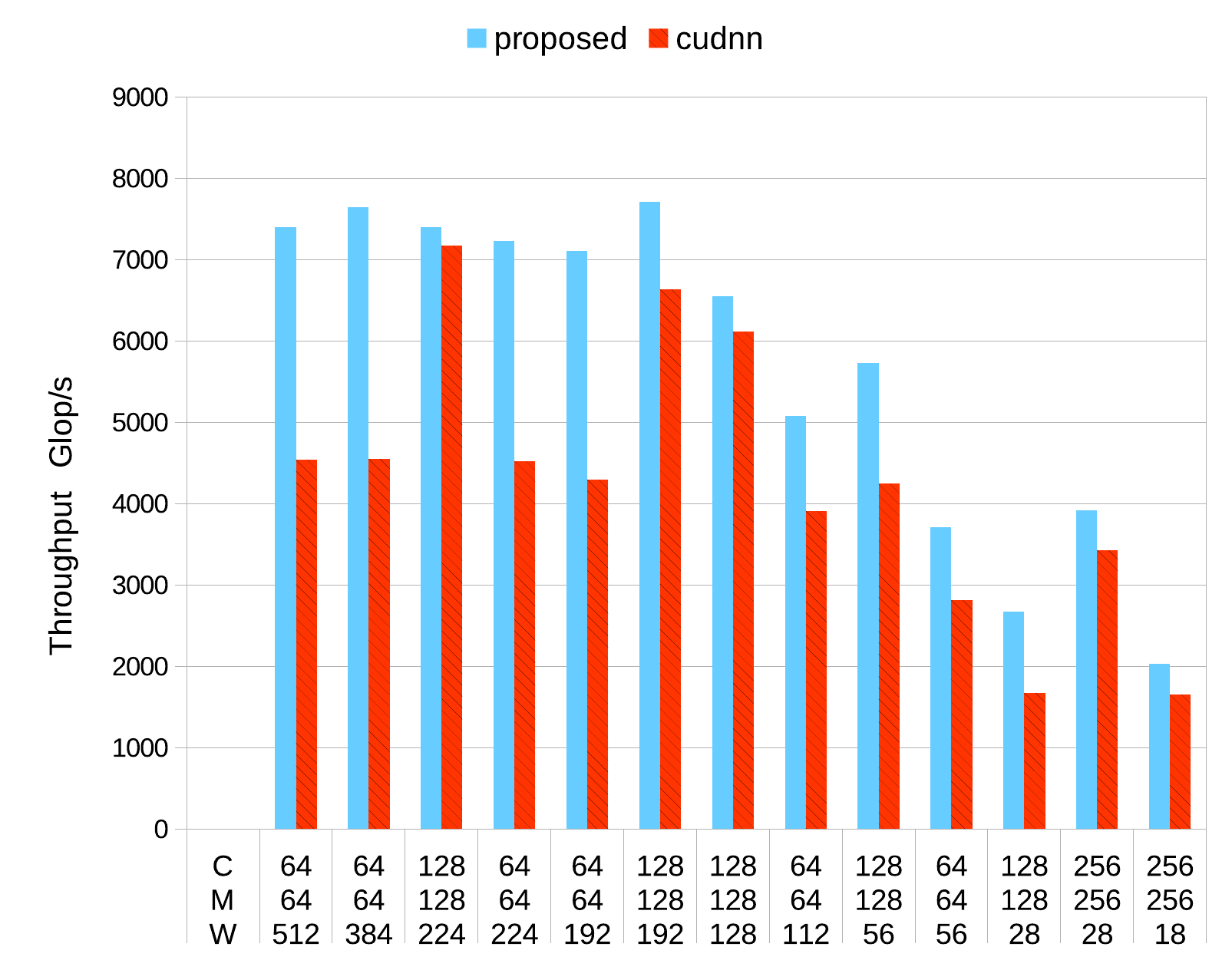}
         {Filter Size = 5}
        \end{center}
      \end{minipage}

    \end{tabular}
    \caption{Performance of the Multi-Channel Convolution Kernel}
    \label{fig:multi-result}
  \end{center}
\end{figure}

In the single-channel convolution, we changed the sample size of the feature
maps from 28 to 1K and the size of the corresponding channels from 512 to
32. The filter size is 1, 3 or 5, which is usually used in many CNN models.
In CUDA programming, by assigning more number of blocks to each $SM$,
the $SMs$ can be kept busy. In our current
implementation, $N_{block} = 2\times{N_{SM}} = 2\times{28}$ blocks are used.
Two blocks are assigned to each $SM$, and 512 threads are assigned to
each block. Thus, the maximum number of registers for each thread is
constrained to 128.
For each tested case, $P$ and $Q$ are decided following the method
described in Section 3.1.
Fig.\ref{fig:single-result} shows the results of the single-channel
convolution. Our method is faster than Cudnn v7.1 in all tested
cases.
The performance gain is 1.5X to 5.6X, and its average is 2.6X.
In the multi-channel convolution, we changed the sample size of the feature
maps from 7 to 512, and the size of the corresponding channels from 64
to 512. The filter size is also 1, 3, or 5.
As discussed in Section 3.2, larger $M'$ is preferable
  for making \textit{data prefetching} more effective. Therefore, we
fixed the segment size $S$ as 32 or 64 bytes, and then
$M'$ and $W'_{x}$ are decided following the method described in Section 3.2.
According to our preliminary evaluation, when $M'=64$ and $W'_{x}=128$,
the performance becomes best, and we used these values for this
comparison.
As shown in Fig.\ref{fig:multi-result}, our method is faster than Cudnn in all
tested cases, and the throughput has been increased by 1.05X to 2X, with an
average increase of 1.39X. In \cite{Op-direct}, a different GPU is used, and a
direct comparision is not possible. However, when $K=3$, our performance is 4X
faster than \cite{Op-direct} on GPU the peak performance of which is 2.4X faster
than that used in \cite{Op-direct}. 
%

We also implemented our two kernels on Maxwell series GPU GTX Titan X, and it
also showed that our performance is faster than Cudnn on the same GPU by 1.3X to
3.7X in the single-channel convolution and 1.08X to 1.8X in the multi-channel
convolution. 

\section{CONCLUSIONS}

In this paper, we proposed two convolution kernels on Pascal series GPUs for
single-channel and multi-channel respectively. For single-channel convolution,
we introduced an effective method of data mapping, which can hide the access
delay of the global memory efficiently. For multi-channel convolution, we
introduced a method that not only guarantees the memory access efficiency, but
also achieves high FMA operation ratio per loaded data. Performance comparison
with the public library Cudnn shows that our approaches are faster in all tested
cases: 1.5X to 5.5X in the single-channel convolution and 1.05X to 2X in the
multi-channel convolution. Our approaches was designed assuming Pascal
architecture, but the performance is also faster than Cudnn on Maxwell
architecture.
This practice shows that our approaches can be applied to the wide range of CNN
models on various GPUs.
In our current implementation, the throughput is still
lower than the theoretical maximum. It means that the convolution kernel still
has the room for improvement, and this is our main future work.


\begin{thebibliography}{15}
  \bibitem{Op-direct} Xiaoming Chen, Jianxu Chen, Danny Z. Chen, Xiaobo Sharon Hu.
    ``Optimizing Memory Efficiency for Convolution Kernels on Kepler GPUs,''In DAC, 2017.
  \bibitem{hand} A. Poznanski, L. Wolf,
   ``Cnn-n-gram for handwriting word recognition,''in: CVPR, 
    pp. 2305-2314, 2016. 
  \bibitem{Deep-CNN} D. Yu, W. Xiong, J. Droppo, A. Stolcke, G. Ye, J. Li, and G. Zweig,
    ``Deep convolutional neural networks with layer-wise context expansion and attention,''in Proc.Interspeech, 2016.
  \bibitem{Chainer} Eddie Bell.
    ``A implementation of squeezenet in chainer.''https://github.com/ejlb/squeezenet-chainer, 2016.
  \bibitem{Mem-hierarchy} X. Mei, X. Chu, 
    ``Dissecting GPU memory hierarchy through microbenchmarking,''IEEE Trans. Parallel Distrib. Syst, 2016.
 \bibitem{VGG} K. Simonyan and A. Zisserman. 
    ``Very deep convolutional networks for large-scale image recognition,''In ICLR, 2015.
  \bibitem{NLP} Zhang, Y. and Wallace, B. 
    ``A sensitivity analysis of (and practitioners guide to) convolutional neural networks for sentence classification.''arXiv:1510.03820, 2015.
  \bibitem{Winograd} A. Lavin. 
    ``Fast algorithms for convolutional neural networks.''arXiv:1509.09308, 2015.
  \bibitem{ResNet} K. He, X. Zhang, et al. 
    ``Deep residual learning for image recognition,''arXiv:1512.03385, 2015.
  \bibitem{Direct} S. Li, Y. Zhang, C. Xiang, and L. Shi. 
    ``Fast Convolution Operations on Many-Core Architectures.''In HPCC, pages 316-323, 2015.
  \bibitem{GoogleNet} Szegedy, C. et al. 
    ``Going deeper with convolutions.''Preprint at http://arxiv.org/abs/1409.4842, 2014. 
  \bibitem{Cudnn} Sharan Chetlur, Cliff Woolley, et al.
    ``cuDNN: Efficient primitives for deep learning.''arXiv:1410.0759, 2014.
  \bibitem{FFT} M. Mathieu, M. Henaff, et al. 
    ``Fast training of convolutional networks through ffts.''In CoRR, 2013.
  \bibitem{Caffe} Jia, Y.
    ``Caffe: An open source convolutional architecture for fast feature embedding.'' http://caffe.berkeleyvision.org/, 2013.
  \bibitem{Alex} Krizhevsky, A., Sutskever, I., and Hinton, G. E.  
    ``ImageNet classification with deep convolutional neural networks,''In NIPS, pp. 1106-1114, 2012.
  \bibitem{Dgemm} Guangming Tan, Linchuan Li, et al.
    ``Fast implementation of DGEMM on Fermi GPU.''In Supercomputing 2011, pages 35:15:11, New York, NY, USA, 2011. ACM.
  \bibitem{Magma} R. Nath, S. Tomov, and J. Dongarra. 
    ``An improved magma gemm for fermi gpus.''Technical Report 227, LAPACK Working Note, 2010.
\end{thebibliography}
\end{document}